\documentclass[a4paper]{jpconf}
\usepackage{graphicx}
\begin{document}
\title{Dilution effects in Ho$_{2-x}$Y$_x$Sn$_2$O$_7$: from the Spin Ice to the single-ion magnet}

\newcommand{\HoSn} {Ho$_2$Sn$_2$O$_7$}
\newcommand{\HoY} {Ho$_{2-x}$Y$_x$Sn$_2$O$_7$}

\author{G. Prando$^{1}$, P. Carretta$^1$, S.R. Giblin$^2$, J. Lago$^1$, S. Pin$^3$, P. Ghigna$^3$}

\address{$^1$ Dip. di Fisica "A.Volta", University of Pavia, 27100 Pavia (Italy)}
\address{$^2$ ISIS Facility - RAL, Chilton Didcot, Oxfordshire OX11 0QX (United Kingdom)}
\address{$^3$ Dip. di Chimica Fisica "M.Rolla", University of Pavia, 27100 Pavia (Italy) }

\ead{giacomo.prando@ghislieri.it}

\begin{abstract}
A study of the modifications of the magnetic properties of \HoY\, upon varying the concentration of diamagnetic
Y$^{3+}$ ions is presented. Magnetization and specific heat measurements show that the Spin Ice ground-state is
only weakly affected by doping for $x\leq 0.3$, even if non-negligible changes in the crystal field at Ho$^{3+}$
occur. In this low doping range $\mu$SR relaxation measurements evidence a modification in the low-temperature
dynamics with respect to the one observed in the pure Spin Ice. For $x\rightarrow 2$, or at high temperature, the
dynamics involve fluctuations among Ho$^{3+}$ crystal field levels which give rise to a characteristic peak in
$^{119}$Sn nuclear spin-lattice relaxation rate. In this doping limit also the changes in Ho$^{3+}$ magnetic
moment suggest a variation of the crystal field parameters.
\end{abstract}

\section{Introduction}

In \HoSn\, the  pyrochlore lattice formed by ferromagnetically coupled Ho$^{3+}$ ions and the  single-ion axial
anisotropy lead to a macroscopically degenerate ground-state where two magnetic moments, out of the four at the
vertexes of each tetrahedron, point out and the other two in. Owing to the analogy with hydrogen atoms arrangement
around oxygen in I$_h$ ice this ground-state has been named Spin Ice \cite{SpinIce}. The stability of this phase
depends on the relative intensity of the effective magnetic coupling among Ho$^{3+}$ magnetic moments, due both to
dipolar and exchange coupling mechanisms, and on the magnitude of single-ion anisotropy \cite{SpinIce}. By
replacing Ho$^{3+}$ with diamagnetic Y$^{3+}$ ions in \HoY , one can progressively reduce the magnetic coupling
still keeping a sizeable single-ion anisotropy and study how the Spin Ice phase is progressively depressed by
magnetic dilution. If one pushes the dilution further to the $x\rightarrow 2$ limit, one ends up with nearly
isolated Ho$^{3+}$ ions, which are possibly still characterized by a $|J=8, M_J= \pm 8\rangle$ two-fold degenerate
ground-state as for $x=0$. This system is quite similar to rare-earth single-molecule magnets, as Tb or HoPc$_2$
\cite{SIM}, which have recently attracted a lot of interest, in view of their possible applicability as logic
units. The comparison of the spin dynamics in inorganic single-ion \cite{Appl} and molecular magnets could allow
to understand the role of the intra and intermolecular vibrational modes in the decoherence mechanisms
\cite{Santini}. Here we will present  magnetization, specific heat, NMR and $\mu$SR measurements in \HoY\, in the
weakly diluted limit $0\leq x\leq 0.3$ and for strongly diluted samples, with $x=1.99$ and $x=1.995$. It will be
shown that Y$^{3+}$ doping can lead to non-negligible changes in the crystal field (CF) parameters. For $x\leq
0.3$, although the macroscopic degeneracy of the Spin Ice ground-state seems to be preserved, a modification in
the low-energy excitations is observed. For $x\rightarrow 2$ the dynamics involve transition among the single-ion
CF levels driven by spin-phonon interactions.

\section{Experimental Results and Discussion}

The experiments to be reported in the following were performed on unoriented powders. The measurements on loose
powders were not always reproducible when magnetic fields ($H$) of a few Tesla were applied, possibly due to a
partial reorientation of the powders. Accordingly, when needed we have sealed the samples in epoxy resin to grant
the reproducibility of the results.

From the study of the $H$ dependence of the magnetization ($M$) at 3 K we have have derived how the expectation
value of Ho$^{3+}$ magnetic moment varies upon increasing $x$ (Fig. 1). For $x=0$ a value close to the one
reported in literature is found \cite{magpowd}. However, as one progressively dilutes \HoSn\, a non-monotonous
behaviour is observed. While for $x\leq 0.3$ the $H$ dependence of $M$ is not sizeably affected by doping, for
$x\rightarrow 2$ significant changes in $M$ vs. $H$ curves can be noticed. These results are different from the
ones reported for Dy$_{2-x}$Y$_x$Ti$_2$O$_7$ \cite{Dy} and suggest that in \HoY\, non negligible variations in the
CF parameters do occur as one progressively increases Y content. In the undoped Spin Ice a significant mixing of
states at different $M_J$ is already present in the first excited states and the modifications in the local
lattice symmetry induced by dilution are expected to affect it. Nevertheless, the similarity in $M$ vs. $H$ curves
at low doping levels indicates that the CF ground-state preserves its strong axial symmetry and hence one could
expect that the Spin Ice ground-state is little affected for $x\leq 0.3$.

\begin{figure}[h!]
\vspace{6cm} \includegraphics{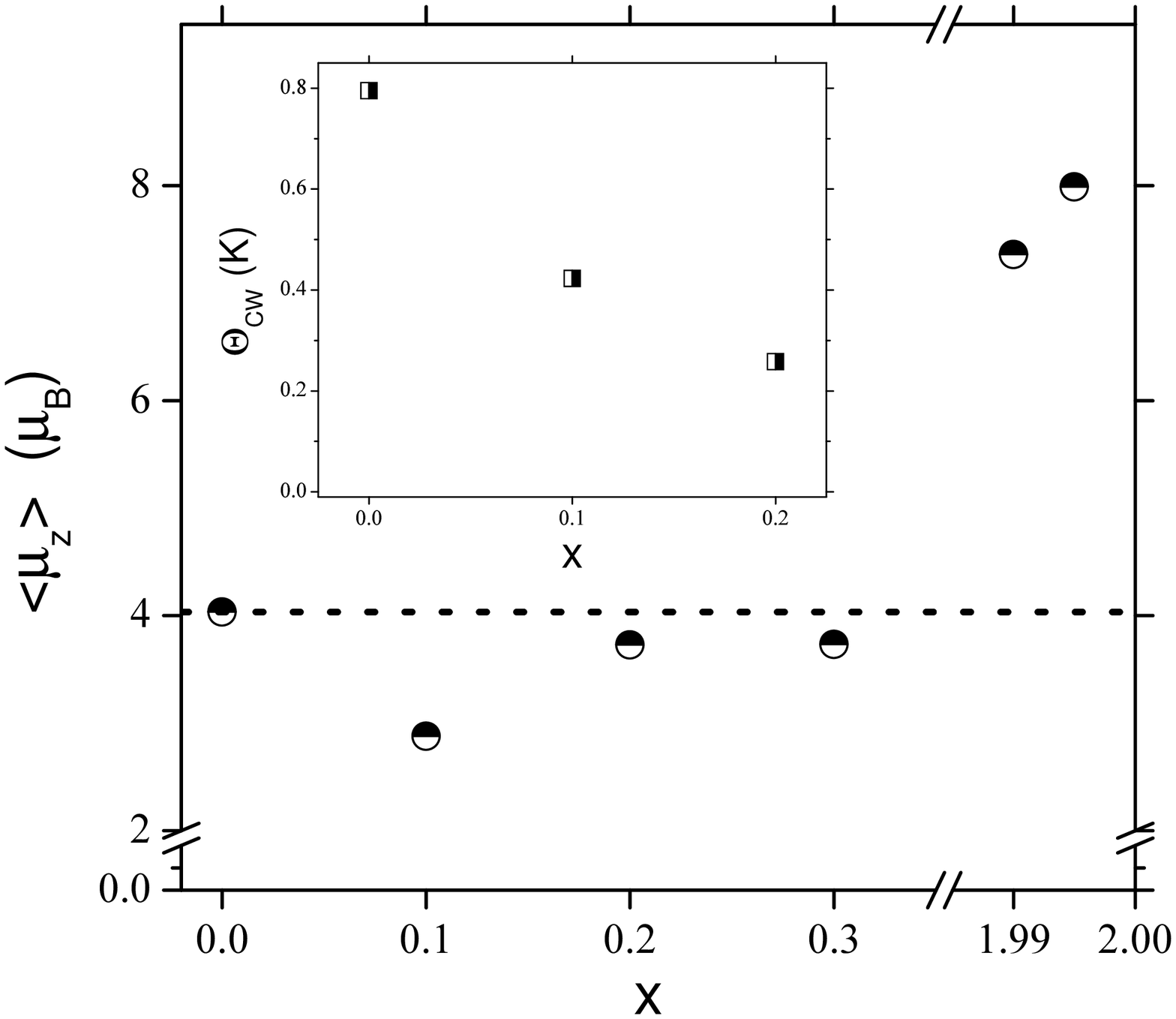} \includegraphics{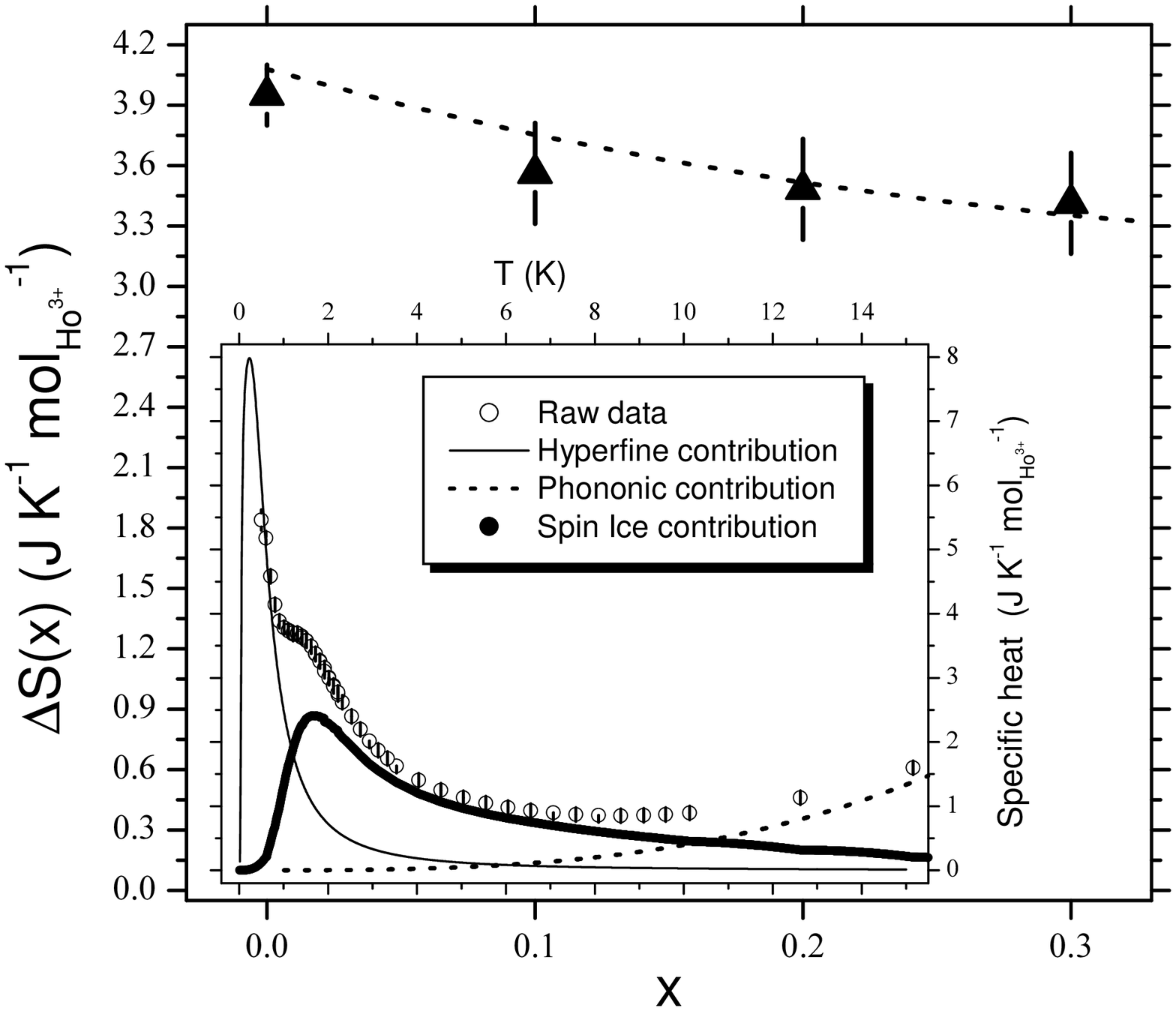} \caption{(left) Expectation value of Ho$^{3+}$ magnetic moment in
\HoY\, as derived from $M$ vs. $H$ curves at $T= 3$ K. In the inset the $x$-dependence of the Curie-Weiss
temperature is shown for $x\leq 0.2$. (right) Entropy variation between $T\rightarrow 0$ and 15 K, as estimated
from specific heat data. The dotted line shows the theoretical behaviour expected by computing Pauling estimate of
the entropy for a diluted system \cite{Sdix}. In the inset the different contributions to the specific heat for
the $x=0$ sample are reported.}
\end{figure}

In fact, from specific heat measurements we have observed that upon increasing the Y content to $x=0.2$ only a
small change in the entropy is present (Fig. 1). This modification is analogous to the one already reported for
Ho$_{2-x}$Y$_x$Ti$_2$O$_7$ \cite{Sdix}, where the entropy variations were accounted for by evaluating the Pauling
entropy for ice with vacancies. It must be noticed that upon increasing $x$ a certain decrease in the Ho hyperfine
contribution to the specific heat is observed. This indicates a decrease in the hyperfine coupling at small doping
levels, which supports the reduction of Ho$^{3+}$ magnetic moment evidenced by magnetization measurements.

Now we turn to the discussion of the modifications in the spin dynamics in the light of $^{119}$Sn NMR and $\mu$SR
measurements. In the strongly diluted limit, or at temperature ($T$) much larger than the interaction among
Ho$^{3+}$ magnetic moments,  the dynamics involves fluctuations among the single-ion CF levels. These transitions
are driven by spin-phonon interaction and cause a decrease in the life-time of the levels. Indirect nuclear
spin-lattice relaxation processes, involving nuclear spin-flips with no change of CF level, are sensitive to these
life-times. In fact, by extending the same description already used to successfully describe nuclear spin-lattice
relaxation rate $1/T_1$ in discrete level systems as the molecular magnets, one can write \cite{Borsa}
\begin{equation}
\frac{1}{T_1}= \frac{\gamma^2\langle\Delta h_{\perp}^2\rangle}{Z} \sum_{m} \frac{e^{-E_m/k_BT}\tau_m}{1+
\omega_L^2\tau_m^2}
\end{equation}
with $\langle\Delta h_{\perp}^2\rangle$ the mean-square amplitude of the hyperfine field fluctuations, $\omega_L$
$^{119}$Sn Larmor frequency, $E_m$ the eigenvalues of the CF levels and $Z$ the corresponding partition function.
It should be observed that $E_m$ depends  on the magnetic field intensity and on its orientation with respect to
the anisotropy axes and that, accordingly, in a powder one has to compute an average of $1/T_1$ over all possible
orientations. $\tau_m$ can be related to the energy difference between the CF levels and to an average spin-phonon
coupling constant $C$ \cite{Villain}. $1/T_1$ vs. $T$ is expected to show a peak  with an intensity decreasing
with the magnetic field. For the $x=1.99$ sample we found a peak in the $50$ K range, while for the $x=0.2$ sample
a peak was observed at much higher temperature, around 130 K (Fig. 2). The shape of these peaks can be
qualitatively reproduced by Eq. 1 if CF parameters different from the ones of Ho$_2$Ti$_2$O$_7$ \cite{CF} are
used. It has to be pointed out that two main factors determine the position of the peak, the CF level spacing and
the spin-phonon coupling constant $C$, and it is not possible to discern from $1/T_1$ data alone how much each one
of them is varying with doping.

\begin{figure}[h!]
\vspace{6cm} \includegraphics{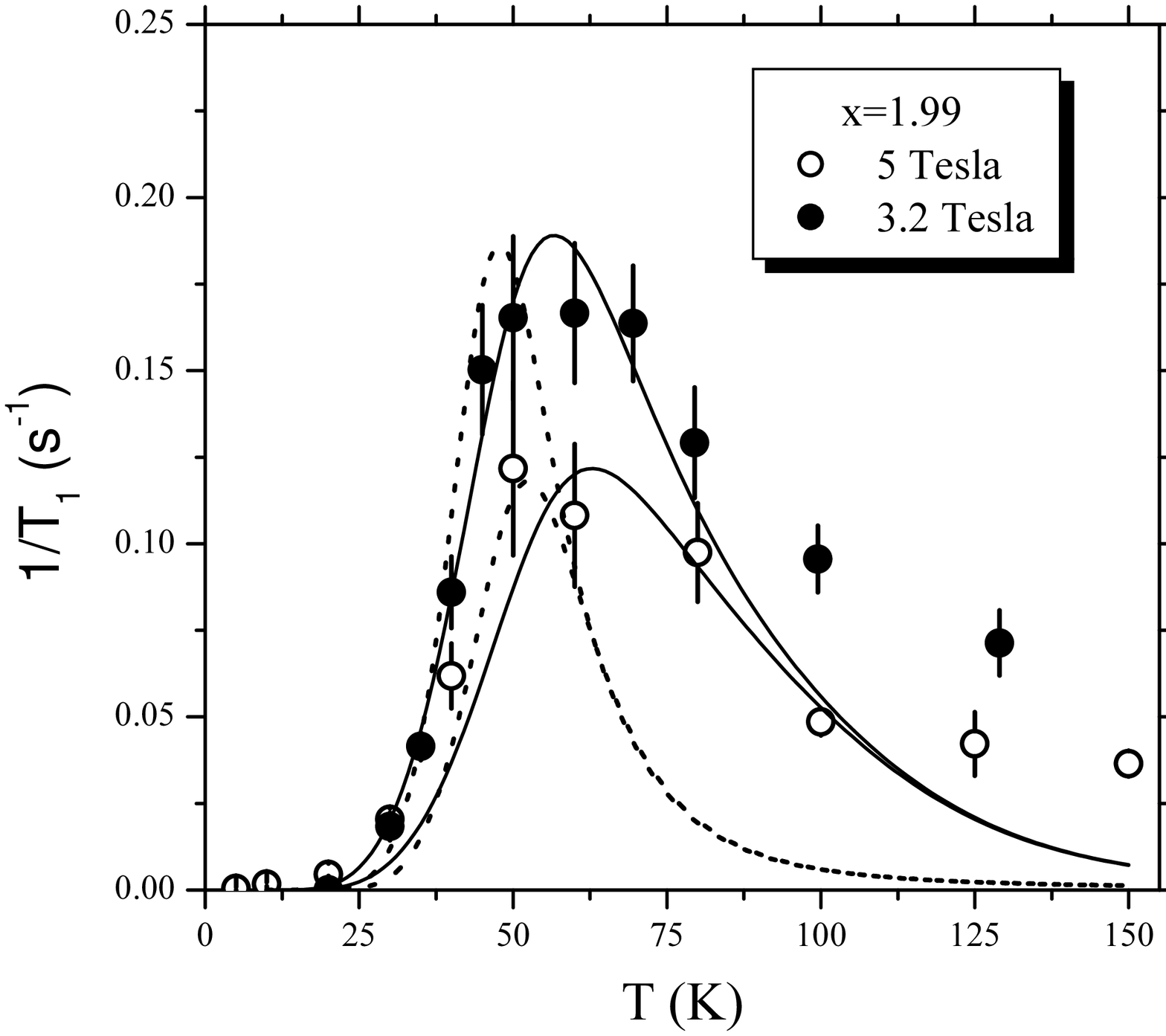} \includegraphics{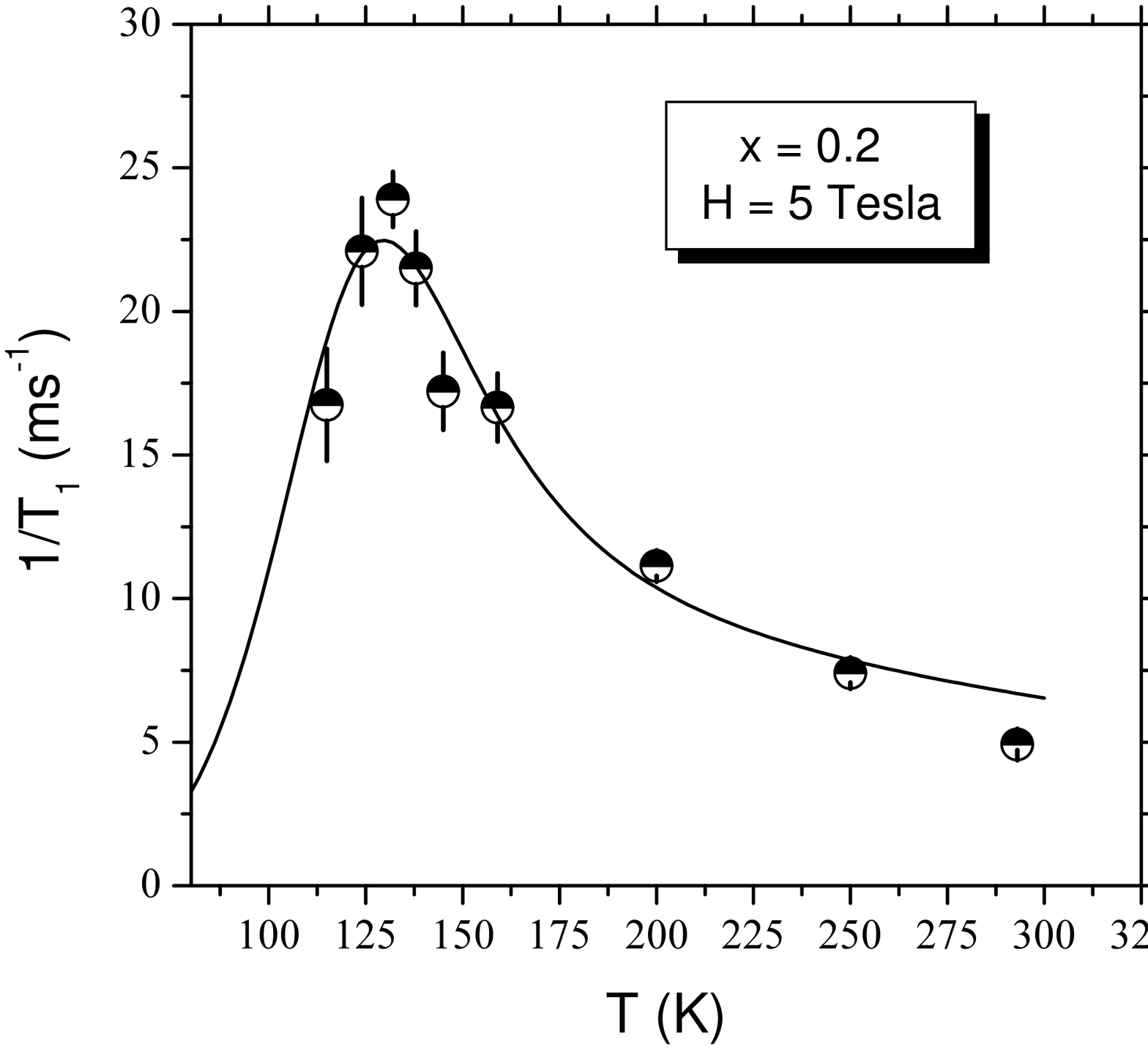} \caption{(left) Temperature dependence of $^{119}$Sn $1/T_1$ in the
$x=1.99$ sample, for $H= 3.2$ and $5$ Tesla.  (right)  Temperature dependence of $^{119}$Sn $1/T_1$ in the $x=0.2$
sample, for $H= 5$ Tesla. The dotted lines show the behaviour according to Eq. 1 by taking the CF parameters of
Ho$_2$Ti$_2$O$_7$ \cite{CF}, while the solid lines refer to the best fits, obtained by modifying the CF levels. }
\end{figure}

At low $T$ $\mu^+$ zero-field relaxation, for $x\leq 0.3$,  is characterized  by an initial fast decrease of the
muon polarization which prevents the observation of 2/3 of the total asymmetry and by a  1/3 tail with a slow
decay, determined by the low-energy excitations within the Spin Ice ground-state. Here we shall describe the
behaviour of the relaxation rate $\lambda$ of the 1/3 tail.  In \HoSn\, $\lambda$ shows a behaviour similar to the
one found in other pyrochlore systems \cite{TbSn}. In fact, $\lambda$ is observed to progressively increase on
cooling owing to the slowing down of the dynamics on approaching the Spin Ice ground-state and then to become
nearly constant below $T\simeq 1$ K, when fluctuations within the degenerate ground-states arise. On the other
hand, in diluted samples one observes a drop in $\lambda$ for $T\rightarrow 0$ (Fig. 3). This decrease can be
ascribed either to a further slowing down or to a lowering in the amplitude of the fluctuations. Both could
originate from the presence of a gap of the order of hundreds of mK between the ground-state and the first excited
states, due to a partial relieve of the degeneracy of the ground-state. It is worth noting that the presence of
spin vacancies is expected to reduce the energy gap required to generate excitations \cite{monopolo}.

\begin{figure}[h!]
\vspace{6.5 cm} \includegraphics{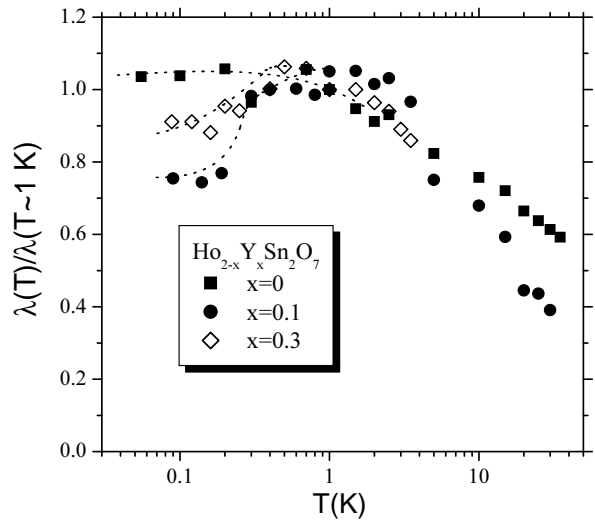} \caption{$T$ dependence of
the muon relaxation rate in zero-field, normalized to its value at $T\simeq 1$ K. $\lambda$ at $T=1$ K are $0.20$,
$0.23$ and $0.24$ $\mu$s$^{-1}$ for the $x=0, 0.1$ and 0.3 samples, respectively. Lines are guide to the eye.}
\end{figure}

\section{Conclusions}

In conclusion it was shown from magnetization and NMR $1/T_1$ measurements that the progressive dilution of \HoY\,
leads to modifications in the Ho$^{3+}$ CF parameters. Still, the Spin Ice ground-state seems to be robust and the
degeneracy of the ground-state is observed to vary little with doping for x up to $0.3$. In this low doping limit,
however, $\mu$SR measurements reveal a novel dynamic which is possibly associated with a partial relieve of the
degeneracy of the ground-state.

\smallskip

\end{document}